\documentclass{svjour3}                     
\smartqed  
\usepackage{graphicx}
\begin{document}

\title{Form factors of heavy-light systems in point-form relativistic quantum mechanics: the Isgur-Wise function  
\thanks{Presented at the 21st European Conference on Few-Body Problems in Physics,
Salamanca, Spain, 30 August - 3 September 2010}
}

\author{
M. G\'omez Rocha \and W. Schweiger 
}


\institute{M. G\'omez Rocha \and W. Schweiger \at
              Institut f\"ur Physik, FB Theoretische Physik, Universit\"at Graz, A-8010 Graz, Austria \\
\email{maria.gomez-rocha@uni-graz.at} \\
\email{wolfgang.schweiger@uni-graz.at}
}

\date{Received: date / Accepted: date}

\maketitle

\begin{abstract}
We investigate electromagnetic and weak form factors of
heavy-light mesons in the context of point-form relativistic
quantum mechanics. To this aim we treat the physical processes
from which such electroweak form factors are extracted by means of
a coupled channel approach which accounts for the dynamics of the
intermediate gauge bosons. It is shown that heavy-quark symmetry
is respected by this formulation. A simple analytical expression
is obtained for the Isgur-Wise function in the heavy-quark limit.
Breaking of heavy-quark symmetry due to realistic values of the
heavy-quark mass are studied numerically.

\end{abstract}
Though the point-form of relativistic quantum dynamics is the
least explored of the three common forms of relativistic dynamics,
it has several properties that makes it well suited for
applications to hadronic physics. Its main characteristics are
that interaction terms (if present) enter all four components of
the 4-momentum operator, whereas the generators of Lorentz
transformations stay free of interactions. As a particular example
we are going to present the calculation of electroweak form
factors of heavy-light mesons within a constituent-quark model.
Since the dependence of matrix elements on the heavy-quark mass is
rather obvious in point-form relativistic quantum mechanics, it is
comparably easy to study the heavy-quark symmetry and its breaking
due to finite masses of the heavy quarks.

Starting point of our investigations are the physical processes
from which such electroweak form factors are extracted, i.e.
elastic electron-meson scattering and the weak decay of
heavy-light mesons. We use a coupled-channel framework in which
the dynamics of the intemediate gauge bosons -- either photon or
W-boson -- is fully taken into account. Poincar\'e invariance is
ensured by emplyoing the Bakamjian-Thomas
construction~\cite{Bakamjian:1953kh}. Its point-form version
amounts to the assumption that the (interacting) 4-momentum
operator $\hat{P}^\mu$ can be factorized into an interacting mass
operator and a free 4-velocity operator
$
 \hat P^\mu =\hat M \hat V^\mu_{\mathrm{free}}\, .
$
It is therefore only necessary to study an eigenvalue problem for
the mass operator.

In case of elastic electron-meson scattering a mass eigenstate
$\hat{M} |\psi \rangle = m |\psi \rangle$ is written as a direct
sum of a quark-antiquark-electron component $|\psi_{Q\bar{q} e}
\rangle$ and a quark-antiquark-electron-photon component $|\psi_{Q
\bar{q} e\gamma} \rangle$. Here we have already assumed that the
quark carries the heavy flavor. The mass eigenvalue equation to be
solved has the form
\begin{eqnarray}\label{eigenvalue:equation}
 \left(\begin{array}{ll} \hat M_{Q\bar{q} e} & \hat K
 \\ \hat K^\dagger &
 \hat M_{Q\bar{q} e\gamma}\end{array}\right)
 \left(\begin{array}{l}  |\psi_{Q\bar{q} e}\rangle
 \\   |\psi_{Q\bar{q} e\gamma}\rangle
 \end{array}\right)   =
 m \left(\begin{array}{l} |\psi_{Q\bar{q} e}\rangle
 \\ |\psi_{Q\bar{q} e\gamma}\rangle
 \end{array}\right)\, ,
\end{eqnarray}
where $M_{Q\bar{q} e}$ and $M_{Q\bar{q} e\gamma}$ consist of a
kinetic term and an instantaneous confining potential between
quark and antiquark, and $\hat K$ is a vertex operator which
accounts for the emission and absorption of a photon by the
electron or (anti)quark. It is determined by the interaction
Lagrangean density of QED~\cite{Klink:2000pp}.

For the calculation of the electromagnetic meson currents and form
factors it is most convenient to apply a Feshbach reduction to the
mass eigenvalue problem and study the optical potential
$\hat{V}_{\mathrm{opt}}(m)=\hat K^\dagger (\hat M_{Q\bar{q}
e\gamma}-m)^{-1}\hat K$. The electromagnetic meson current
$J^\mu(\vec{k}^\prime_M;\vec{k}_M)$ can then be extracted from the
invariant 1-$\gamma$-exchange amplitude which is essentially given
by on-shell matrix elements of the optical potential. These have
the structure
\begin{eqnarray}\label{eq:contract} \mathcal{M}_{1\gamma}
(\vec{k}_e^\prime, \mu_e^\prime;\vec{k}_e, \mu_e) &\propto&
\langle V^\prime; \vec{k}_e^\prime, \mu_e^\prime; \vec{k}_M^\prime
\vert \hat{V}_{\mathrm{opt}}(m) \vert V; \vec{k}_e, \mu_e;
\vec{k}_M\rangle_{\mathrm{on-shell}}\nonumber\\ & \propto & V^0
\delta^3(\vec{V}-\vec{V}^\prime)\frac{ j_\mu(\vec{k}_e^\prime,
\mu_e^\prime; \vec{k}_e, \mu_e)
J^\mu(\vec{k}_M^\prime;\vec{k}_M)}{(k_e^\prime-k_e)^2}\, .
\end{eqnarray}
$\vert V; \vec{k}^{(\prime)}_e, \mu^{(\prime)}_e;
\vec{k}^{(\prime)}_M\rangle$ are, so called, \lq\lq velocity
states\rq\rq\ that specify the state of a system by the overall
velocity and the center-of-mass momenta and canonical spins of its
components~\cite{Klink:1998zz}. In our case $\vec{k}^{(\prime)}_M$
is the momentum of the confined $q$-$\bar{q}$ subsystem with the
quantum numbers of the heavy-light meson. \lq\lq On-shell" means
that $m=k_e^0+k_M^0=k_e^{\prime \,0}+k_M^{\prime\, 0}$ and
$k_e^0=k_e^{\prime \,0}$, $k_M^0=k_M^{\prime \,0}$. A detailed
derivation of Eq.~(\ref{eq:contract}) and the explicit expression
for the meson current $J^\mu(\vec{k}_M^\prime;\vec{k}_M)$ can be
found in Ref.~\cite{Gomez:2010}.

If we are dealing with a pseudoscalar meson its electromagnetic
current should be of the form
$J^\mu(\vec{k}_M^\prime;\vec{k}_M)=({k}_M^\prime+{k}_M)^\mu
F(Q^2)$, which allows us to identify the electromagnetic form
factor of the meson uniquely. It is, however, known that the
Bakamjian-Thomas construction, that we are using, provides wrong
cluster properties~\cite{Keister:1991sb}. As a consequence, the
hadronic current $J^\mu(\vec{k}_M^\prime;\vec{k}_M)$ which we
extract from Eq.~(\ref{eq:contract}) exhibits a slight dependence
on the electron momenta $k_e$ and $k_e^\prime$. Fortunately this
dependence vanishes rather quickly with increasing invariant mass
$m$ of the electron-meson system and thus also in the heavy-quark
limit ($m_Q=m_M \rightarrow \infty$, $m_{\bar{q}}/m_Q \rightarrow
0$). This limit has to be taken in such a way that $v ^\prime\cdot
v  = 1 + Q^2/2 m_M^2$ stays constant. The function of
$(v^\prime\cdot v)$ that is obtained from $F(Q^2)$ by taking the
heavy-quark limit is the famous Isgur-Wise
function~\cite{Isgur:1989vq}. In our case it takes on a rather
simple analytical form:
\begin{eqnarray}\label{eq:xiem}
\xi(v'\cdot v)= \lim_{m_Q\rightarrow \infty} F(Q^2) &=&
\sqrt{\frac{1}{2(1+v\cdot v')}} \sum_{\mu' \mu} \int d^3 \tilde
k'_{\bar{q}}\, \sqrt{\frac{\tilde \omega_{\bar q}}{\tilde
\omega'_{\bar q}}} \, \psi_{\mathrm{out}}(\vec{\tilde k}'_{\bar
q})\, \psi_{\mathrm{in}}(\vec{\tilde k}_{\bar q}) \nonumber\\
& & \times D^{1/2}_{\mu' \mu}\left[
R^{-1}_W\left(\frac{\tilde{k}_{\bar q}}{m_{\bar q}},B(v)\right)
R_W\left( \frac{\tilde{k}'_{\bar q}}{m_{\bar q}},B(v')\right)
\right]\, .
\end{eqnarray}
It is just an integral over incoming and outgoing wave functions,
a Wigner-rotation factor and kinematical factors. The tildes in
the integral indicate that the corresponding quantities are given
in the $Q$-$\bar{q}$ rest system. In accordance with heavy-quark
symmetry $\xi(v'\cdot v)$ does not depend on the heavy-quark mass.

\begin{figure}
\begin{center}
  \includegraphics[width=0.45\textwidth]{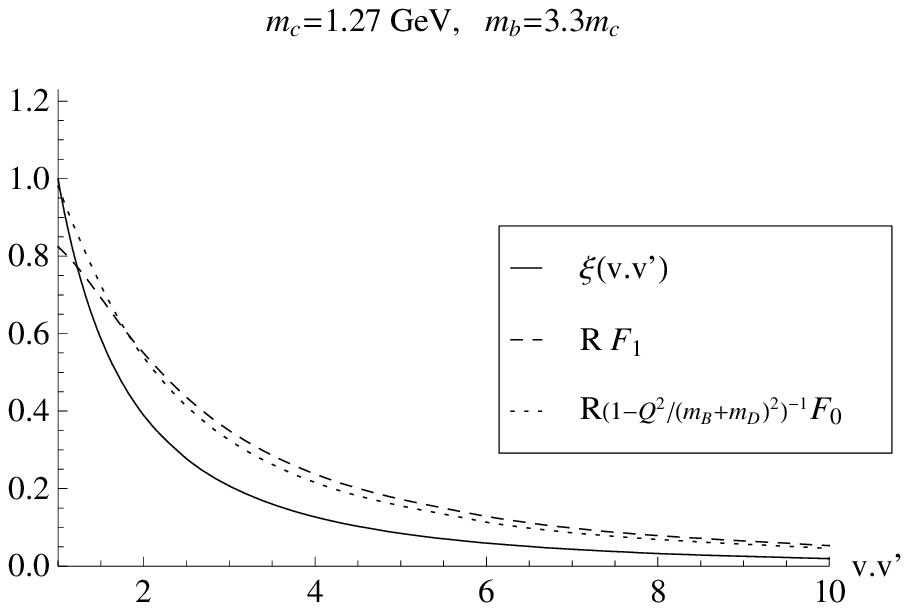}
  \includegraphics[width=0.45\textwidth]{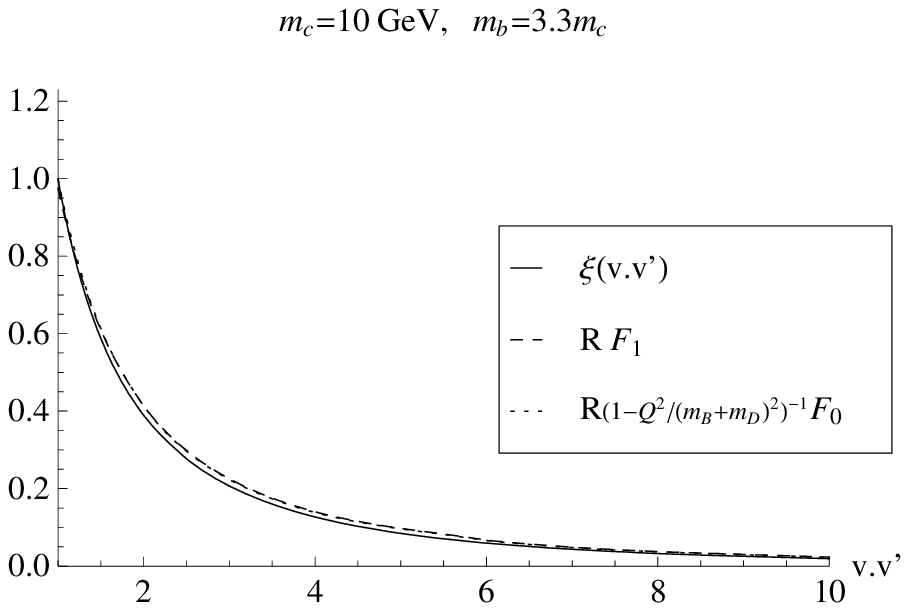}
\end{center}
\caption{Weak decay form factors (multiplied with appropriate
kinematical factors) for the process $B^-\rightarrow D^0e^-\bar
\nu$ for finite $m_Q$ in comparison with the Isgur-Wise function.
For the $Q$-$\bar{q}$ bound-state wave function we have taken a
Gaussian with the same oscillator parameter ($a=0.55$~GeV) and
light-quark mass ($m_{u, d}=0.25$~GeV) as in
Ref.~\cite{Cheng:1996if}. \vspace{-0.3cm}}
\label{fig:2}       
\end{figure}

Heavy-quark symmetry allows us also to relate the electromagnetic
form factor of a pseudoscalar heavy-light meson to its weak decay
form factors for heavy-to-heavy flavor transitions (like, e.g.,
$\bar{B}\rightarrow D^{(\ast)} \ell \bar{\nu}$). In the
heavy-quark limit the electromagnetic form factor and the weak
decay form factors (modulo kinematical factors) should lead to
only one Isgur-Wise function~\cite{Neubert:1993mb}. If we apply
our coupled channel framework to semileptonic decays of
pseudoscalar heavy-light mesons, identify the decay form factors
from the optical potential and take the heavy-quark limit we end
up again with $\xi(v'\cdot v)$. These investigations show that
heavy-quark (flavor) symmetry is recovered in the heavy-quark
limit within our relativistic coupled channel approach.

It is also interesting to study the breaking of heavy-quark
symmetry caused by finite values of the heavy-quark mass. This is
done in Fig.~\ref{fig:2} for the two weak decay form factors
$F_0(v'\cdot v)$ and $F_1(v'\cdot v)$ that show up in the
semileptonic $B^-\rightarrow D^0e^-\bar \nu$ decay. If heavy-quark
symmetry was perfect $RF_1$ and $R(1-q^2/(m_B+m_D)^2)^{-1}F_0$
(with $R=2\sqrt{m_B m_D}/(m_B+m_D)$) should coincide with the
Isgur-Wise function $\xi(v'\cdot v)$ (see
Ref.\cite{Neubert:1993mb}). What we rather observe is that the
physical values of the $b$- and $c$-quark mass give rise to a
considerable breaking of heavy-quark symmetry (left plot). Here we
have not even taken into account a (heavy) flavor dependence of
the meson wave functions. If both masses were about a factor of 10
larger heavy-quark symmetry would nearly hold (right plot).

A more comprehensive study of heavy-light systems along the lines
presented here, including the discussion of (heavy-quark) spin
symmetry,  can be found in Ref.~\cite{Gomez:2010}.

\vspace{-0.4cm}

\begin{acknowledgements}
M. G\'omez Rocha acknowledges the support of the \lq\lq Fonds zur
F\"orderung der wissenschaftlichen Forschung in \"Osterreich" (FWF
DK W1203-N16).
\end{acknowledgements}

\end{document}